\documentclass[10pt,letterpaper,twocolumn]{article} %% two column, final layout
\pdfoutput=1

\usepackage{ol}
\usepackage{hyperref}

\newcommand{\um}{$\mu$m}
\newcommand{\minusone}{$^{-1}$}

\newcommand{\omegapl}{\omega_\text{pl}}
\newcommand{\eperp}{\epsilon_\perp}
\newcommand{\epar}{\epsilon_\|}
\newcommand{\neff}{n_\text{eff}}
\newcommand{\absval}[1]{\left|\,#1\right|}

\usepackage{amssymb}
\usepackage{amsmath}
\usepackage{amsfonts}
\usepackage{latexsym}
\usepackage{graphicx}%
\usepackage{braket}

\begin{document}

\twocolumn[

\title{Non-magnetic negative-refraction systems for terahertz and far-infrared frequencies }% Force line breaks with \\

\author{Leonid V. Alekseyev$^1$, Viktor A. Podolskiy$^2$ and Evgenii E. Narimanov$^1$}
\affiliation{${}^1$ Electrical and Computer Engineering Department, Purdue University, West Lafayette, IN 47907 \\
${}^2$ Physics Department, University of Massachusetts, Lowell, MA 01854}

% \maketitle % in revtex must come after abstract

\begin{abstract}
We demonstrate that homogeneous naturally-occurring  materials can form non-magnetic  negative refractive index systems, and
 present specific realizations of the proposed approach for the THz and far-IR
 frequencies. The proposed structure operates away from resonance, thereby promising the capacity for low-loss devices.
\end{abstract}

]
% \maketitle % in revtex must come after abstract

% In the past several years, negative refractive index materials have transitioned from the domain of wishful thinking by the theorists into experimentally realizable implementations.

% In the past several years, increasing body of research on negative refractive index materials resulted in their transition from
% an electromagnetic curiosity of questionable utility to being prime candidates for novel optical devices.

Following the initial proposal by Veselago in
1968~\cite{veselago}, negative refraction materials spent over 30
years as a forlorn curiosity before being resurrected with renewed
interest from both theoretical and experimental groups.  Within
the last several years it was realized that these materials (known
also as left-handed materials) possess unusual properties, some of
which were not recognized at the time of their conceptions.  These
properties include resonant enhancement of evanescent fields,
potentially enabling near-perfect imaging below the diffraction
limit and leading to a new class of optical
devices~\cite{parimi_nature_03}, as well as nontrivial behavior in
the nonlinear regime~\cite{gabitov}.  Despite initial controversy
over the realizability of negative index materials (NIMs),
successful proof of principle demonstrations have been
accomplished~\cite{shelby_science_01, parimi_nature_03,
shalaev_OL_2005,park_PRB_2006,karlsruhe_slow_light}.

Existing designs for left-handed materials rely on achieving
overlapping dipolar and magnetic resonances~\cite{smith_prl_00,
parazzoli_prl_03}, or using photonic crystals near the
bandgap~\cite{notomi_prb_00, parimi_nature_03}.  Both of these
approaches necessitate complicated 3D patterning of the medium
with microstructured periodic arrays.  Fabrication of such
structures presents significant challenges even for GHz
applications, while manufacturing systems for higher frequencies
becomes harder still. Furthermore, near-resonant operational
losses impose severe limitations on the imaging
resolution~\cite{nssl}.

As an alternative to periodic systems, a waveguide-based
implementation of a NIM was proposed~\cite{nmlhm}, which obviates
the need for negative magnetic permeability and does not require
periodic patterning. This approach circumvents major manufacturing
obstacles to achieving NIM behavior at terahertz or optical
frequencies.

To achieve this behavior, the waveguide material must possess
characteristics of a uniaxial medium with a significant
anisotropy. Furthermore, this anisotropy must ensure that
$\epsilon_\perp$ (the component of $\epsilon$ transverse to the
planar waveguide) is negative, while $\epsilon_\|$ (in-plane
component) remains positive.  TM modes  in such waveguide undergo
negative refraction in the waveguide plane, and propagate with
negative phase velocity~\cite{nmlhm}.

One of the key aspects in designing such an NIM system is
selecting the material for the waveguide's anisotropic core.
Several options have been proposed for the core material, in
particular, nanostructured composites in a dielectric host and
quantum well structures~\cite{nmlhm}. While being within the grasp
of existing technology, the fabrication of such systems remains
highly challenging.

In this Letter we present an alternative approach to non-magnetic
NIMs for THz and far-infrared domains based on naturally occurring
materials with large dielectric anisotropy. In particular, we
discuss the possibility of negative refraction in a
waveguide-based system at approximately 20 \um, 58 \um\ and 255
\um\ using, respectively, sapphire, bismuth, or triglycine sulfate
in the waveguide core.  We focus on monocrystalline bismuth as an
attractive option for manufacturing the NIM waveguide core thanks
to its large anisotropy and availability of samples with high
purity.

% (having transverse permittivity $\epsilon_\perp$ and in-plane permittivity
% $\epsilon_\|$),

In a planar waveguide with anisotropic dielectric core the wave
vector components $k_z$ and $k_y$ are governed by the dispersion
relation
\begin{equation}\label{eq:dr}
k_z^2+k_y^2=\epsilon \, \nu \frac{\omega^2}{c^2},
\end{equation}
with $\nu = \left(1-\kappa^2 c^2/\epsilon_\| \, \omega^2\right)$
%\frac{\kappa^2 c^2}{\epsilon_\| \omega^2}\right),$
where $\omega$ is the frequency of light,
$\epsilon=\epsilon_\perp(\epsilon_\|)$ for the TM(TE) modes,
$\kappa$ is the transverse mode parameter, and $k_z$ and $k_y$ lie
in the waveguide plane. For perfectly conducting waveguide walls
(a good approximation for silver and other metals at THz and
far-IR frequencies~\cite{Wangberg2005}), $\kappa=m \pi /d$, where
$m$ is an integer and $d$ is the thickness of the
waveguide~\cite{nmlhm}. The effective refractive index for
propagating waveguide modes in this system is given by $\neff^2 =
\epsilon \, \nu$. To support propagating modes, $\epsilon$ and
$\nu$ must have the same sign. The case $\epsilon
> 0$, $\nu > 0$ is typically realized in an isotropic planar waveguide
operating above cut-off.  However, in the case $\epsilon < 0$,
$\nu < 0$  negative refraction
occurs~\cite{nmlhm}, with refractive index given by
\begin{equation}\label{eq:neff}
\neff = -\sqrt{\eperp \, \nu}.
\end{equation} Note that if $\kappa$ is regarded as the transverse wave vector
component, Eq.(\ref{eq:dr}) can be rewritten as a hyperbolic
dispersion relation
$k^2_\perp/\epar-k^2_\|/\absval{\eperp}=\omega^2/c^2$, identified
in Ref.~[\onlinecite{joannopoulos_prb_02}] with the onset of
negative refraction.

\begin{figure*}[htbp]
\centering
\includegraphics[width=14cm]{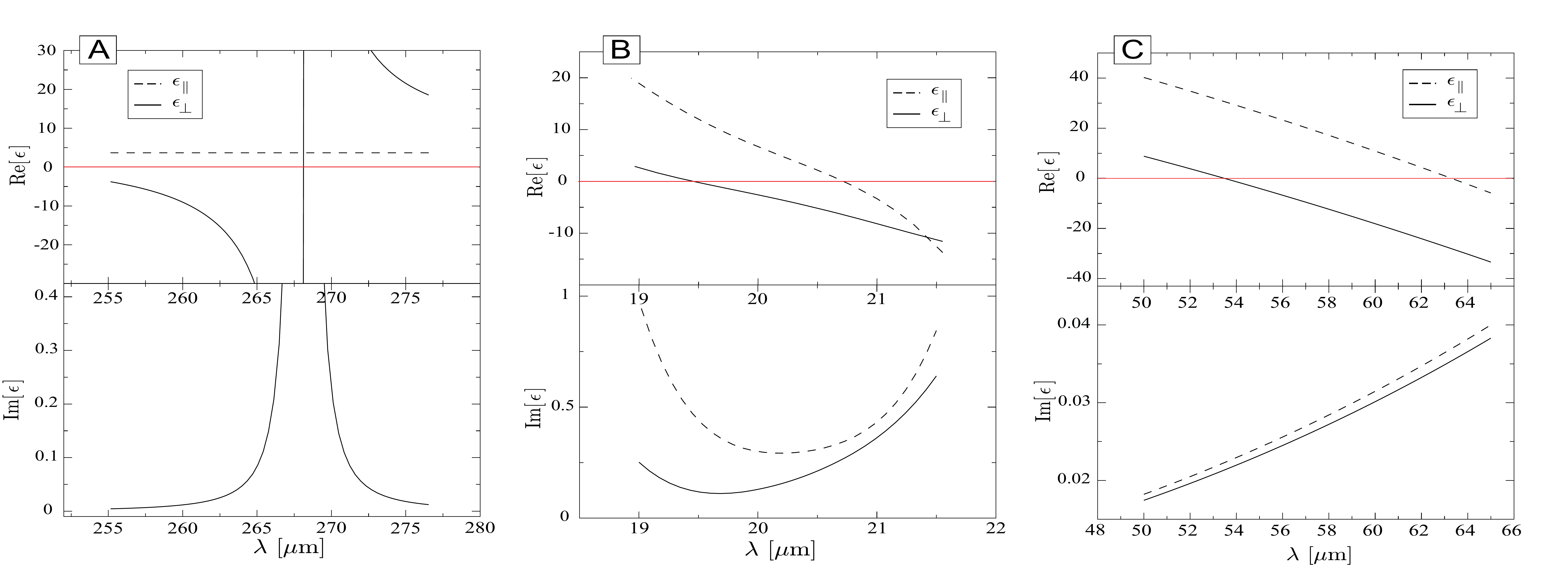}
\caption{(A): The real (top panel) and the imaginary (bottom
panel) parts of the dielectric function of TGS; the monoclininc
$C_2$ axis is along the ``perpendicular" ($\perp$) direction.
(B):~Same for sapphire; the crystallographic $c$ axis is along the
``perpendicular" ($\perp$) direction. (C):~Same for
monocrystalline bismuth. } \label{fig:dielFns}
\end{figure*}

As follows from the definition of $\nu$, the propagation with
negative phase velocity occurs only for TM modes when
$\epsilon_\parallel > 0, \epsilon_\perp < 0$. This behavior is
observed in a number of materials where structural anisotropy
strongly affects the dielectric response.

One example of such materials is triglycine sulfate (TGS), a
compound widely used in fabricating infrared photodetectors.
Spectroscopic studies of the crystal at low temperature have shown
that phonon modes polarized parallel to the crystal's monoclinic
$C_2$ axis significantly differ in frequency from phonons
transverse to the axis.  This results in a large anisotropy in the
dielectric tensor along these directions.  In particular,
dielectric response for the field polarized {\em along} the $C_2$
axis features a resonance at 268~\um, which is absent if the
incident field is polarized {\em transverse} to the $C_2$
axis~\cite{tgs_00}. Dielectric function $\epsilon_\perp$ in the
vicinity of this resonance can be fitted with the Lorentz-Drude
model~\cite{tgs_98}, while $\epsilon_\perp$ in this region can be
taken approximately constant~\cite{tgs_00,dumelow_05}.
Lorentz-Drude model parameters from Ref.~[\onlinecite{tgs_98}]
were used to construct Fig.~\ref{fig:dielFns}(a).  As is evident
from the figure, $\epsilon_\perp < 0$, while $\epsilon_\| > 0$ in
the region 250 $\le \lambda \le$ 268~\um.  Furthermore, the
imaginary part of $\epsilon$ becomes small away from the
resonance, minimizing absorption.  A TGS-filled waveguide with
$C_2$ axis oriented perpendicular to the waveguide plane would
support negative index propagation, while suffering from minimal
propagation losses (Im[$\epsilon$]$\sim$10$^{-3}$ at 250~\um).

Whereas the phonon anisotropy of TGS exists in the low-THz domain,
for other materials, it may occur in a different spectral band. In
particular, the strong anisotropy of the dielectric response of
sapphire (Al$_2$O$_3$) is also due to excitation of different
phonon modes (polarized either parallel or perpendicular to the
$c$ axis of the rhombohedral structure), but occurs around 20~\um.
Fig.~\ref{fig:dielFns}(b) shows
experimentally-determined~\cite{sapphire} $\epsilon_\parallel$ and
$\epsilon_\perp$ as functions of frequency.  As with TGS, a region
of $\epsilon_\perp < 0$, $\epsilon_\| > 0$ is evident in the
experimental data.  This potentially enables a sapphire-based
waveguide NIM (with the $c$ axis of sapphire core perpendicular to
the waveguide plane). Note that the minimum of the material
absorption occurs in the frequency range of interest.

Anisotropic phonon excitations are not the only mechanism that can
lead to strong dielectric anisotropy.  Bismuth, a Group V
semimetal with rhombohedral lattice and trigonal symmetry,
exhibits such anisotropy due to a substantial difference in its
electron effective masses along different directions in the
crystal.

In the frequency region of interest, the spectral dependence of
the electric permittivity of Bismuth can be adequately described
by the Drude model,

\begin{equation}\label{eq:drude}
\epsilon = \epsilon_L \left(1-\frac{\omegapl^2}{\omega^2 + i
\omega \tau^{-1}}\right),
\end{equation}
with $\epsilon_L$ the lattice permittivity, $\omegapl=N
e^2/\epsilon_L m_\text{eff}$ the plasma frequency, and $\tau$ the
relaxation time.  These parameters are known from interferometric
and reflectance studies of Bi samples.  In particular, plasma
frequency of pure Bismuth at 4~K was measured to be
158~cm\minusone\ for the incident $E$-field polarized
perpendicular to the trigonal axis, and 186~cm\minusone\ for the
field polarized parallel to the axis~\cite{bb_58}.  These values
are in agreement with other experiments~\cite{bb_60, ke_74}. The
lattice dielectric constant $\epsilon_L$ for the field
perpendicular to the trigonal axis was found to be
110$\pm$10~cm$^{-1}$~\cite{ke_74}, in reasonable agreement with
Ref.~\onlinecite{bb_60}.  For polarization parallel to the
trigonal axis, $\epsilon_L$=76~\cite{edelman_76}.

\begin{figure}[!thp]
\centering
\includegraphics[width=8.4cm]{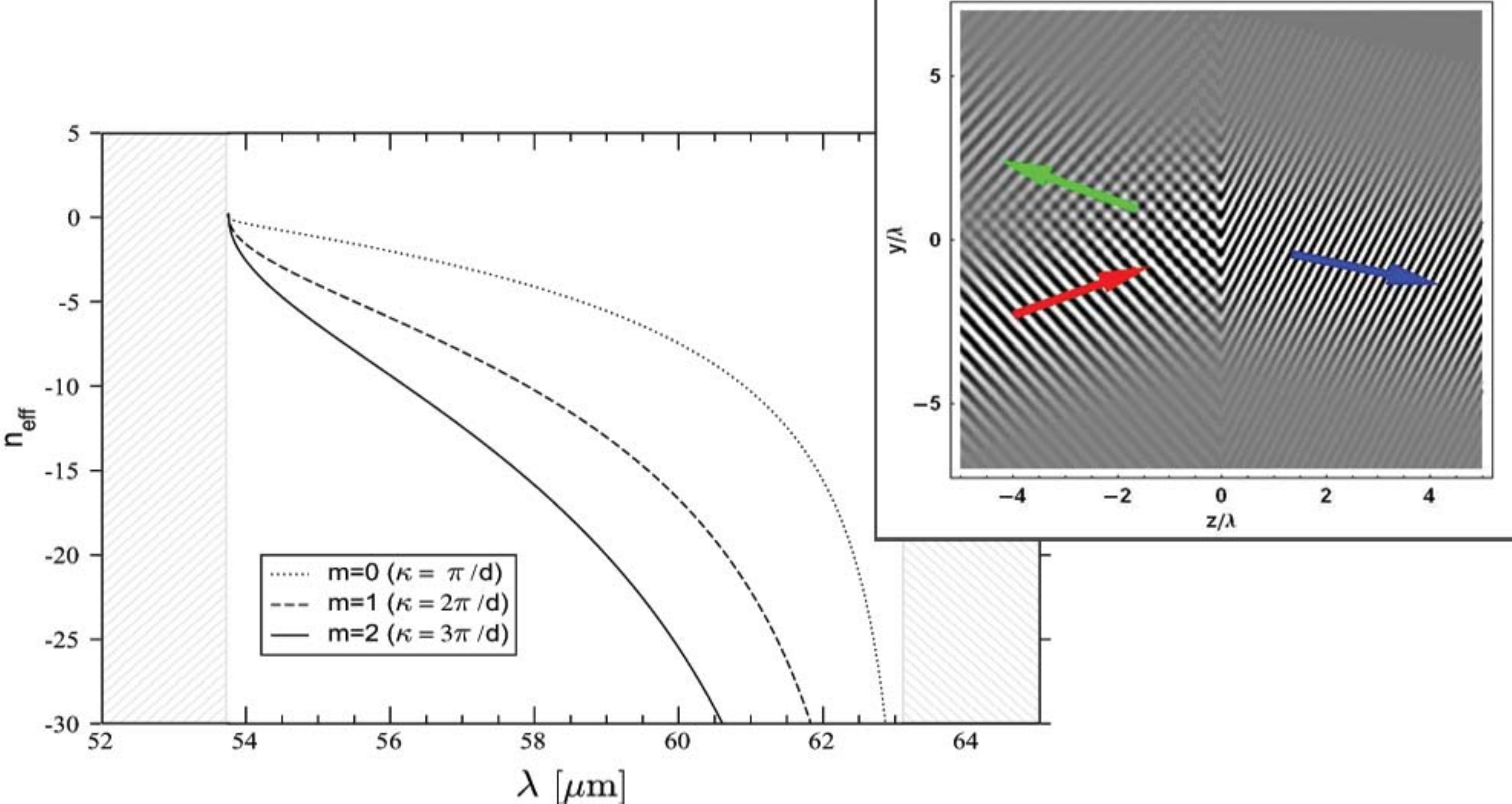}
\caption{ The  effective refractive index of the three lowest
order modes in a bismuth waveguide.  Inset: numerical simulation
showing the refraction of a beam within a metallic waveguide (of
thickness $d = 4.5$~\um) at an interface between an isotropic
dielectric with $\epsilon = 55$ and monocrystalline bismuth;
$\lambda = 61$~\um.} \label{fig:neff}
\end{figure}

There can be substantial variation in the relaxation time $\tau$
depending on the purity of the sample.  We take
$\tau$=0.1~ns~\cite{ke_74}, however, this is a conservative
estimate; for low temperatures, relaxation times over an order of
magnitude greater have been reported as far back as
1975~\cite{edelman_76}. Even with $\tau$=0.1~ns, the typical ratio
of imaginary and real parts of the dielectric function in Bi is on
the order of 0.1\% in the frequency interval of interest, which
enables many imaging and transmission applications~\cite{nssl}. It
should also be noted that high-quality single-crystal films as
thin as 1~$\mu$m, with the trigonal axis ($C_3$) oriented
perpendicular to the film plane, have been
reported~\cite{yang_science_99}, thereby essentially solving the
technological issues in fabricating the proposed negative index
device.

Fig.~\ref{fig:dielFns}(c) shows the behavior of real and imaginary
components of $\epsilon$ for Bi based on Eq.(\ref{eq:drude}). The
most prominent feature of these plots, the transition from
$\epsilon > 0$ to $\epsilon < 0$, is determined by the highly
anisotropic plasma frequency. This anisotropy creates a window
between $\lambda=53.7$~\um\ and 63.2~\um\ where $\epsilon < 0$ for
the $E$-field along the $C_3$ axis, while $\epsilon > 0$ for $E$
transverse to $C_3$. The existence of such 10~\um\ window was
confirmed by direct measurement~\cite{gg_76}.

%
% $(\vecb{E} \perp \vecb{C_3})$
%

To allow for left-handed propagation in this frequency interval,
Bi should be integrated into the core of a planar waveguide, with
the $C_3$ axis oriented in the transverse direction.  In
Fig.~\ref{fig:neff} we examine the behavior of the effective
refractive index $\neff$ [Eq.(\ref{eq:neff})] for the proposed
subcritical ($d < \lambda/2$) waveguide structure with bismuth
core.  Note that negative effective index is possible for all
modes over the entire ($\eperp<0\, ,\,\epar>0$) range. Negative
refraction behavior of our system was further confirmed by a
numerical calculation of the electric field incident on the Bi
waveguide.  The results of this calculation are presented in
Fig.~\ref{fig:neff} (inset). We assume a TM wave with a gaussian
profile, mode-matched into the Bi waveguide in the transverse
direction by e.g. propagating the beam from a metallic waveguide
of the same thickness, filled with a regular dielectric.  One can
clearly see the negative refraction at the boundary.  Furthermore,
it is evident that attenuation of the transmitted wave is weak, as
expected from low values of the imaginary part of the dielectric
constant [Fig.~\ref{fig:dielFns}(c)].

In addition, this calculation shows that the NIM waveguide remains
transparent despite the fact that transverse dimension of the
waveguide is much smaller than the wavelength, which indicates
strong confinement of the field within the core (since the
cladding is assumed to be perfectly conducting). This behavior is
not found in a subwavelength dielectric waveguide, where much of
the field spreads into the cladding, or a subwavelength metallic
waveguide, which does not support propagating modes.  Such strong
field confinement may find applications in photonic structures and
nonlinear optics~\cite{funnels}.

In conclusion, we have proposed a novel negative refraction system
for several wavelengths from low-THz to far-IR.  Our approach is
non-magnetic, avoids the use of periodic patterning, utilizes
naturally occurring materials, and promises the capacity for
low-loss devices.

This work was partially supported by NSF grants DMR-0134736,
ECS-0400615, the Princeton Institute for the Science and
Technology of Materials (PRISM), GRF-OSU, and ACS-PRF.

\bibliography{bi_paper}% Produces the bibliography via BibTeX.

\end{document}